\documentclass[twocolumn,secnumarabic,amssymb, nobibnotes, aps, prb, superscriptaddress,showkeys]{revtex4-1}
\usepackage{graphicx}
\usepackage{dcolumn}
\usepackage{bm}
\usepackage{hyperref}
\usepackage{caption}
\usepackage{float}
\usepackage{url}
\usepackage{mdsymbol}
\usepackage{amsmath, amsthm, amssymb}
\usepackage{svg}
\usepackage{lipsum}
\usepackage{mathrsfs}
\usepackage{eucal}
\usepackage{verbatim}
\usepackage{float}
\usepackage{mathtools}
\usepackage{gensymb}

\begin{document}
	
\title{MagNet: machine learning enhanced three-dimensional magnetic reconstruction}

\author{Boyao Lyu}
\thanks{These authors contributed equally to this work}
\affiliation{Anhui Province Key Laboratory of Condensed Matter Physics at Extreme Conditions, High Magnetic Field Laboratory of Chinese Academy of Sciences, Hefei, 230031, China}
\affiliation{University of Science and Technology of China, Hefei, 230031, China}
\author{Shihua Zhao}
\thanks{These authors contributed equally to this work}
\affiliation{Department of Physics and Astronomy, University of New Hampshire, Durham, New Hampshire 03824, USA}
\affiliation{Department of Physics, The City College of New York, New York, NY 10031, USA}
\affiliation{Physics Program, Graduate Center of the City University of New York, New York, NY 10016, USA}
\author{Yibo Zhang}
\affiliation{Department of Physics and Astronomy, University of New Hampshire, Durham, New Hampshire 03824, USA}
\affiliation{Department of Chemistry, University of New Hampshire, Durham, New Hampshire 03824, USA}
\author{Weiwei Wang}
\affiliation{Institutes of Physical Science and Information Technology, Anhui University, Hefei 230601, China}	
\author{Haifeng Du}
\affiliation{Anhui Province Key Laboratory of Condensed Matter Physics at Extreme Conditions, High Magnetic Field Laboratory of Chinese Academy of Sciences, Hefei, 230031, China}		
\author{Jiadong Zang}
\email{Jiadong.Zang@unh.edu}
\affiliation{Department of Physics and Astronomy, University of New Hampshire, Durham, New Hampshire 03824, USA}
\affiliation{Materials Science Program, University of New Hampshire, Durham, New Hampshire 03824, USA}
%

\begin{abstract}
Three-dimensional (3D) magnetic reconstruction is vital to the study of novel magnetic materials for 3D spintronics.
Vector field electron tomography (VFET) is a major in house tool to achieve that.
However, conventional VFET reconstruction exhibits significant artefacts due to the unavoidable presence of missing wedges. 
%
%
In this article, we propose a deep-learning enhanced VFET method to address this issue.
%
%
A magnetic textures library is built by micromagnetic simulations.
MagNet, an U-shaped convolutional neural network, is trained and tested with dataset generated from the library. 
We demonstrate that MagNet outperforms conventional VFET under missing wedge. 
Quality of reconstructed magnetic induction fields is significantly improved. 
\end{abstract}
\keywords{Vector field electron tomography, deep learning, micromagnetism}

\maketitle
\section{Introduction}
\label{intro}

Recent studies of novel magnetic materials with topological textures, 
such as skyrmionic families~\cite{back20202020,gobel2021beyond,yu2011near,zheng2017direct}, become an important driving force in spintronics to develop next generation nano-electronic devices~\cite{fert2017magnetic,hirohata2020review}.
In addition to two-dimensional (2D) topological textures, their three-dimensional (3D) counterparts are emergent, such as the skyrmion bundle and magnetic hopfion~\cite{tang2021magnetic, liu2020three}.
3D magnetic textures are prominent due to their potentially larger volume-density and novel dynamics.
However, imaging a 3D magnetic configuration is a major obstacle. 
Most existing magnetic imaging tools such as Kerr microscopy~\cite{scheinfein1990180}, magnetic force microscopy~\cite{schwarz2008magnetic}, and spin-polarized scanning tunneling microscopy~\cite{wiesendanger2009spin} can only resolve magnetic configurations on the 2D surface of a sample.
Recent advances in 3D magnetic imaging have been made.
Neutron scattering~\cite{rossat1991investigation, mook1993polarized,muhlbauer2019magnetic}, magnetic X-ray dichroism~\cite{thole1992x, van1999magnetic, donnelly2017three, hierro2020revealing} and Lorentz transmission electron microscopy (LTEM)~\cite{de2009recent} can probe the internal magnetic structure of a sample. 
Compared to neutron scattering and X-ray dichroism, LTEM and its derivatives can achieve sub-Angstrom~\cite{kisielowski2008detection} resolution without accelerating particles with a synchrotron.  
It is thus attractive to enable LTEM-based 3D magnetic reconstructions.

3D vector field electron tomography (VFET), {\it i.e.} 3D magnetic reconstruction from electron phase shifts retrieved from electron holography (EH)~\cite{Midgley2009EH} or transport of intensity (TIE) equation~\cite{volkov2002new},is a relatively new but fast developing 3D magnetic imaging technique. 
Compared to LTEM, phase retrieval in EH significantly elevates the spatial resolution of the imaging.
Since its earliest proposal by Lai $et$ $al.$ in 1994~\cite{lai1994three}, 
the theoretical foundation of VFET has been established~\cite{stolojan2001three, lade2005electron, phatak2008vector}.
Once clean electron phase shifts of two orthogonal and complete tilt series are collected, two components of the magnetic induction field $\mathbf{B}$ can be reconstructed separately by the central slicing theorem in scalar tomography.  
The third component of $\mathbf{B}$ can then be calculated by the constraint $\mathbf{\nabla \cdot B} = 0$. 
%
Thus conventional analytical algorithms, such as weighted backprojection method (WBP) and regridding reconstruction method (Gridrec) can be directly extended to VFET~\cite{phatak2015iterative}. 
However, in real experiments, there are many sources of inevitable errors during electron phase shifts collection, such as noise, sparsity, misalignment, and missing wedge.
Those errors thus lead to significant inevitable artefacts.
%
Iterative algorithms such as algebraic reconstruction technique (ART) and simultaneous iterative reconstruction technique(SIRT), as the second generation of reconstruction algorithms, show the capability of working with data with missing-wedge problem and sparse sampling problem~\cite{phatak2015iterative}.  
Recent advances of iterative algorithms, such as model based iterative reconstruction (MBIR)~\cite{prabhat20173d, mohan2018model}, incorporate with physical knowledge and geometrical information of the sample as prior knowledge and can reconstruct the three components simultaneously.
But iteratively minimizing a cost function has to pay a price of eight times longer run-time compared to conventional analytical methods~\cite{mohan2018model}. 
%

\begin{figure*}[htb!]
	\centering
	\includegraphics[width=0.9\textwidth]{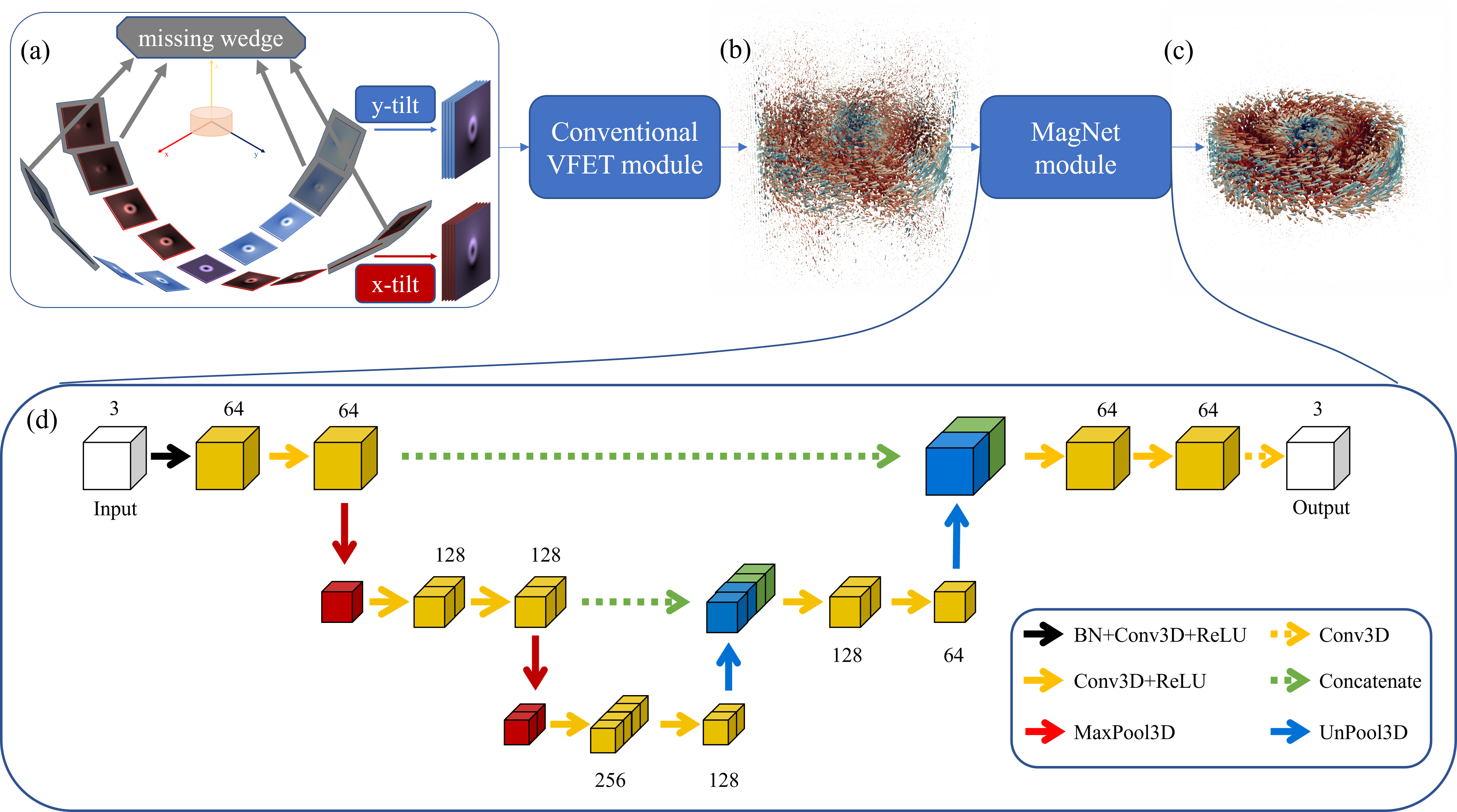}		
	\caption{Workflow of MagNet.} 
	{(a) Limited-angle phase shifts $\phi$ of $x$ and $y$ tilt series. 
	Gray images indicate phase shifts within missing wedge.
	(b) Conventional VFET reconstructed $\mathbf{B}_{in}$. 
	(c) MagNet enhanced $\mathbf{B}_{out}$. 
	(d) The neural network architecture of MagNet. 
	Every convolutional layer is labeled with it's output channel number.}
	\label{fig:flow}
\end{figure*}	

With the development of machine learning techniques, deep learning tomography (DLT) is emergent as the third generation reconstruction algorithm.
%
%
Model with Unet~\cite{ronneberger2015} architecture has already shown its capability in removing artefacts in limited-angle tomography~\cite{gu2017multi}. 
Instead of building an end-to-end DLT algorithm, combining conventional reconstruction with deep learning is an alternative approach to improve the reconstruction results~\cite{adler2018learned}.

In this article, we are focusing on solving the missing-wedge problem in VFET by a DLT algorithm. 
By attaching an Unet architecture machine learning model to conventional VFET, we build a data-driven DLT algorithm that can work end-to-end from phase shifts to $\mathbf{B}$. 
%

We will start section \ref{background} with the theoretical background and dataflow of our reconstruction model followed by a description of our MagNet architecture. 
Details about training and testing samples generation as well as the creation of 3D magnetic textures library will also be discussed in this section. 
In section \ref{results}, reconstruction results will be shown with comparison to the conventional method. Model performance at different missing-wedge conditions are also discussed there.
Conclusions and outlook will be discussed in section \ref{conclusion}.

\maketitle
\section{Theoretical Background, Network architecture, Data Library and Model training}
\label{background}

\begin{figure*}[htb!]
	\includegraphics[width=0.9\textwidth]{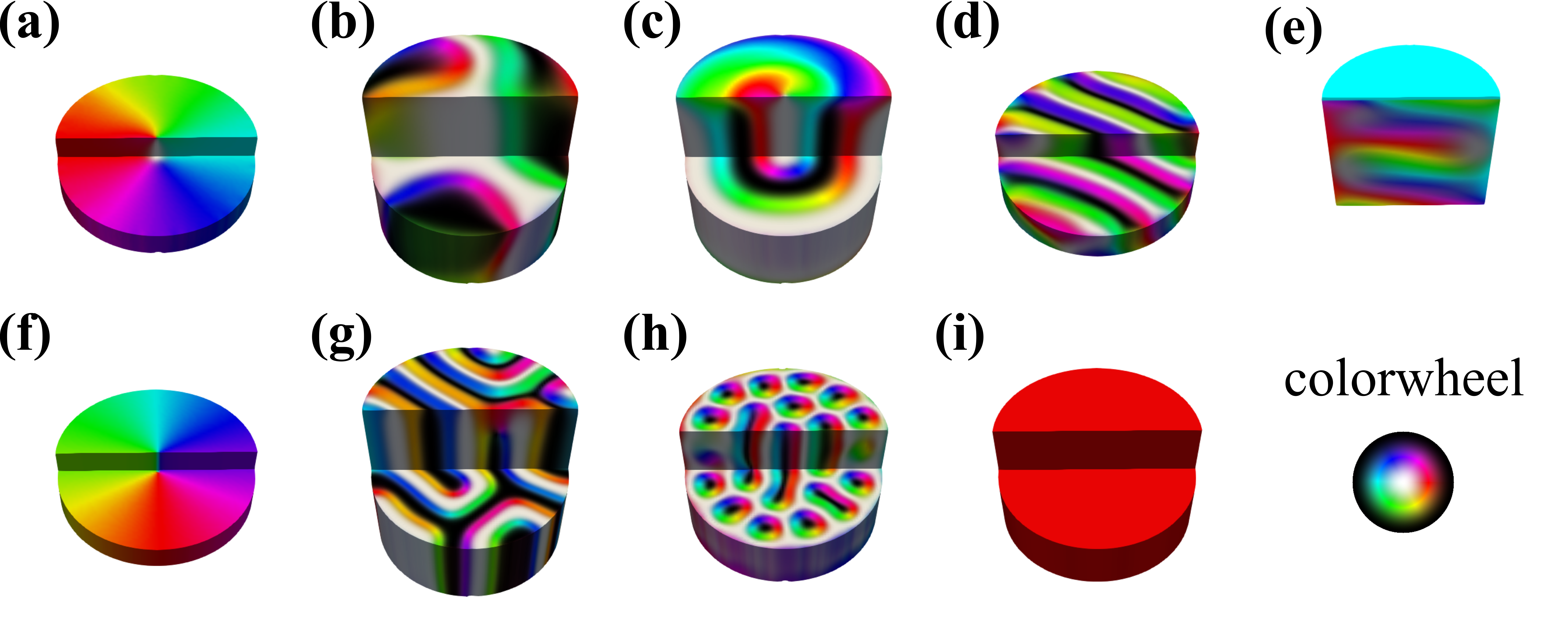}
	\caption{Selected magnetic textures from the library. The view point is from $x+$ direction.
	The $x>0, z>0$ quarter is clipped out to show the interior structure.
		(a) vortex (b) cylindrical domains
		(c) skyrmion (d) spin helix (e)conical
		(f) N\'eel vortex (g)N\'eel domain
		(h) skyrmion lattice (i)single domain.}
	\label{fig:examples}
\end{figure*}

Our MagNet framework is shown in Figure \ref{fig:flow}.
Limited angle phase shifts first enter a VFET module. 
At the presence of missing wedge, the VFET module gives a defective reconstructed magnetic induction field noted as $\mathbf{B_{in}}$. 
$\mathbf{B_{in}}$ is then fed to the Unet module. 
Unet module outputs an enhanced reconstruction result noted as $\mathbf{B_{out}}$.  

In the VFET module, the projected components of the magnetic induction are the gradient of the phase along orthogonal directions~\cite{phatak2008vector}, 
\begin{equation}
	\begin{aligned}
		\partial_y \phi (x,y) &= -\frac{e}{h}\int B_x (x,y,z) dz\\
		\partial_x \phi (x,y) &= \frac{e}{h}\int B_y (x,y,z) dz
	\end{aligned}
\end{equation}
where 
$\phi (x,y) = - \frac{e}{h} \int A_z (x,y,z) dz$
is the phase shift, $e$ is electron charge, and $h$ is the Planck constant. 
Because $B_x$ component is invariant under the rotation about $x$-axis and $B_y$ component is invariant under the rotation about $y$-axis, the reconstruction of $B_x$ and $B_y$ can be simplified as two scalar tomography. 
In this paper, this tomography is achieved by a simple k-space bilinear interpolation~\cite{brynolfsson2010using}.
And the third component is calculated by the constraint $\nabla \cdot \mathbf{B} = 0$.
The Unet module inherits the 3D-Unet skeleton~\cite{Ronneberger2016}.
Convolution blocks are used to extract features. 
The unpool layer is concatenated with skip layer to return the same dimension as the input.
Details of our Unet architecture are shown in Figure \ref{fig:flow} (e).

The geometry of magnetic textures are set as cylinders with radius of 40 pixels and thickness varying from 10 to 80 pixels.
The majority of magnetic structures in our library are generated from micromagnetic software JuMag.jl~\cite{jumag} by setting various simulating parameters and initial states. 
Magnetic textures are then manually selected from simulation results to make sure that sample balance and diversity are taken care of. 
Those micromagnetic structures include vortices, skyrmions, skyrmion lattices, spin helix, conical structures, cylindrical domains, and N\'eel domain structures.
Additionally, manually settled N\'eel vortices and single domain structures are also added to the library.
Representative samples of each category are shown in Figure \ref{fig:examples}.
%
It shows the diversity of our library despite the library size is only 210. 
The whole library is divided into a training set with 150 samples, and a testing set with 60 samples.
Phase shifts for each sample with maximum tilt angle of $\pm 30 \degree$, $\pm 45 \degree$, $\pm 60 \degree$, and $\pm 90 \degree$ are used as the input of the VFET module under different missing wedge conditions. 
The tilt angle step is fixed at $5 \degree$.

The mean squared error (MSE) of the pixel-wise value difference between the output $\mathbf{B_{out}}$ and the ground truth $\mathbf{B_{ref}}$ is chosen as the loss function:
\begin{equation}
	E_{rms}=
	\frac{1}{N^3}
	\sum_{ijk} {[\mathbf{B_{ref}} (i,j,k) - \mathbf{B_{out}} (i,j,k ) ]^2}
\end{equation}
where $N = 100$ is the field-of-view (FOV) pixels of the input and the output.
Adam optimizer with learning rate $lr = 1 \times 10^{-4}$ is used to update the weights and bias during training. 
The Unet module is built with Keras framework~\cite{chollet2015keras}. 
The training was carried out on a single NVIDIA Tesla V100-SXM2-16GB graphic card.
One epoch takes about 329 seconds.
It takes 100 epochs to get MagNet model ready to use.
%

\begin{figure*}[htb]
	\includegraphics[width=0.9\textwidth]{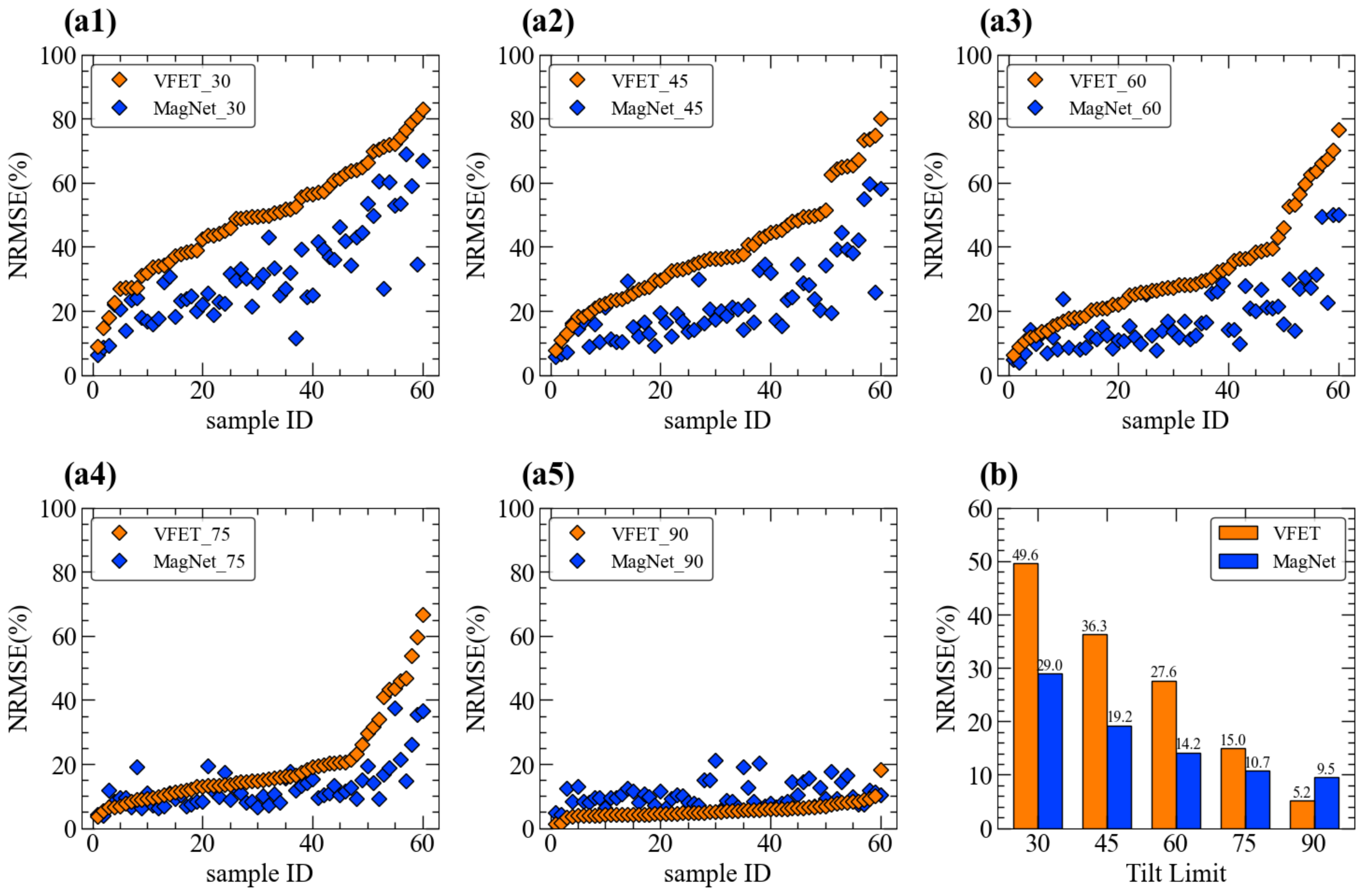}
	\caption{(a1)-(a5) NRMSE distribution of testing data for tilt limit of 
	30$\degree$, 45$\degree$, 60$\degree$, 75$\degree$, and 90$\degree$.
		(b) Median value of each plot.  }
	\label{NRMSE}
\end{figure*}
\maketitle
\section{results and discussion}	
\label{results}
We evaluate the quality of real-space field reconstruction by normalized root mean square error (NRMSE), defined as:
\begin{equation}
	\text{NRMSE} (x,x') = \left\{
	\frac{ \sum_{ijk}^V  \left[x(i,j,k)-x'(i,j,k)\right]^2 }
	{V  \max ({x^2(i,j,k)})}
	\right\}^{\frac{1}{2}}
\end{equation}
where V is the count of summed pixels.
Because the shapes of testing samples are not the same, only pixels inside the material body are taken into account.
The VFET result from $\pm x$ degree tilt series are noted as VFET\_x, and the corresponding MagNet prediction are noted as MagNet\_x.

Figure \ref{NRMSE} (a1)-(a5) show the NRMSE distribution of VFET and MagNet when the maximum tilt angles are 30, 45, 60, 75 and 90, respectively.
The reconstruction is performed in noise-free situation.
All samples are sorted by their NRMSEs of VFET from low to high.
The NRMSEs of VFET are shown in orange dots, and the NRMSEs of MagNet are shown in blue dots.
For example, an orange dot above a blue dot indicates VFET has a higher NRMSE than MagNet for a certain testing sample.	
It can be found that the blue dots are generally below the orange dots in low tilt limit situation  (30$\degree$, 45$\degree$, and 60$\degree$).
%
The advantage becomes less as the tilt limit rises to 75$\degree$.
%
And when the tilt series is complete, MagNet has a higher NRMSE than VFET.
The NRMSE distribution also shows that the influence of tilt limit varies from sample to sample.
For example, the NRMSEs of $\text{VFET}\_60$ are distributed between $10\%$ and $75\%$.
For such a wide distribution, we use the median value for evaluating the reconstruction quality, as shown in Figure\ref{NRMSE} (b).
%
The median NRMSEs of VFET are $49.6\%$, $36.3\%$, $27.6\%$, $15.0\%$ and $5.2\%$ when the tilt limits are 30$\degree$, 45$\degree$, 60$\degree$, 75$\degree$ and 90$\degree$. 
And those of MagNet are $29.0\%$, $19.2\%$, $14.2\%$, $10.7\%$ and $9.5\%$, respectively.
This metric also shows that MagNet has a better performance except when the tilt series is complete.
%

\begin{center}
	\begin{figure}[htb]
		\includegraphics[width=0.45\textwidth]{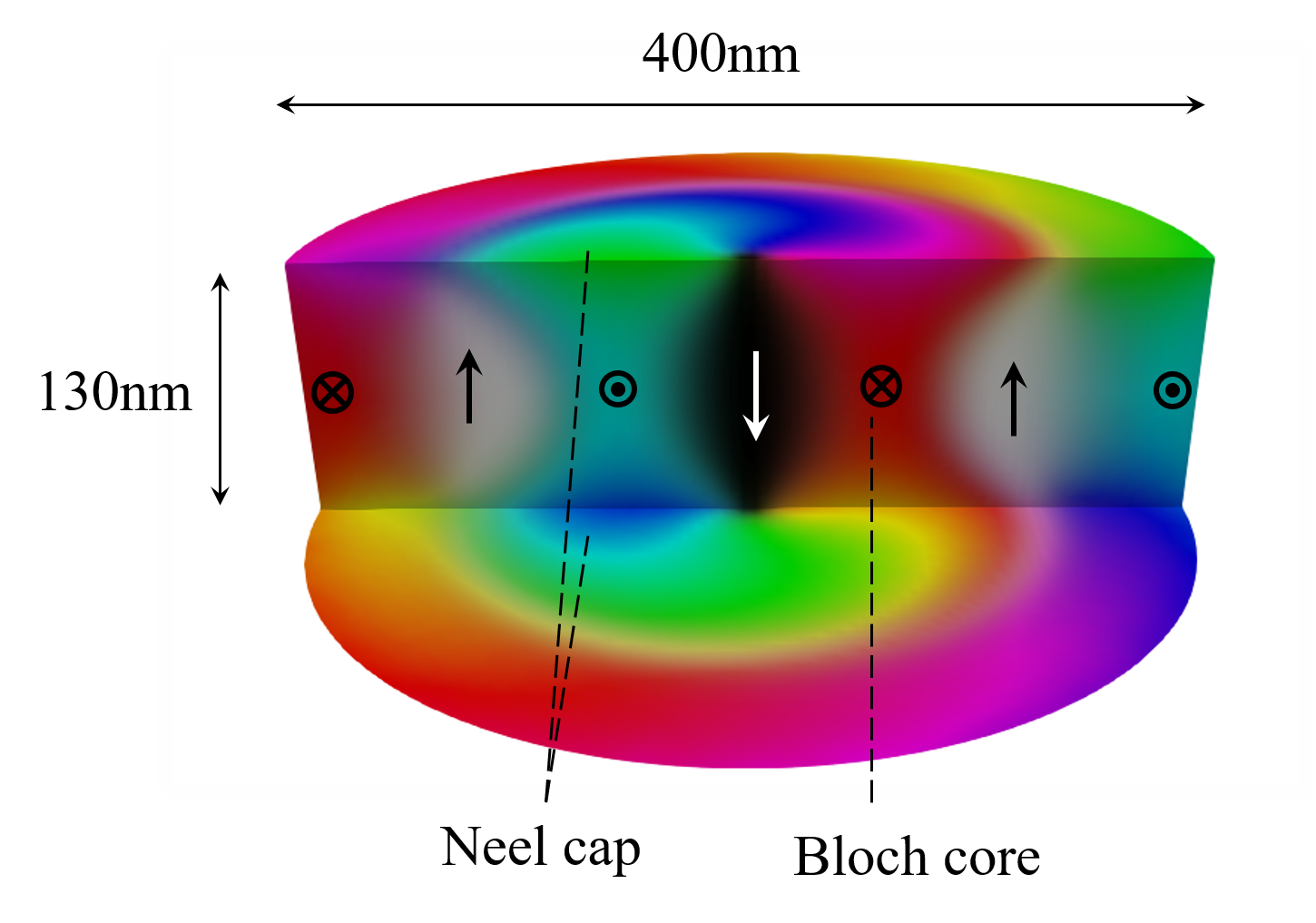}
		\caption{Magnetic texture of the demo example. The view point is from $x+$ direction.}
		\label{mag}
	\end{figure}
\end{center}
%

\begin{figure*}[htb]
	\includegraphics[width=0.8\textwidth]{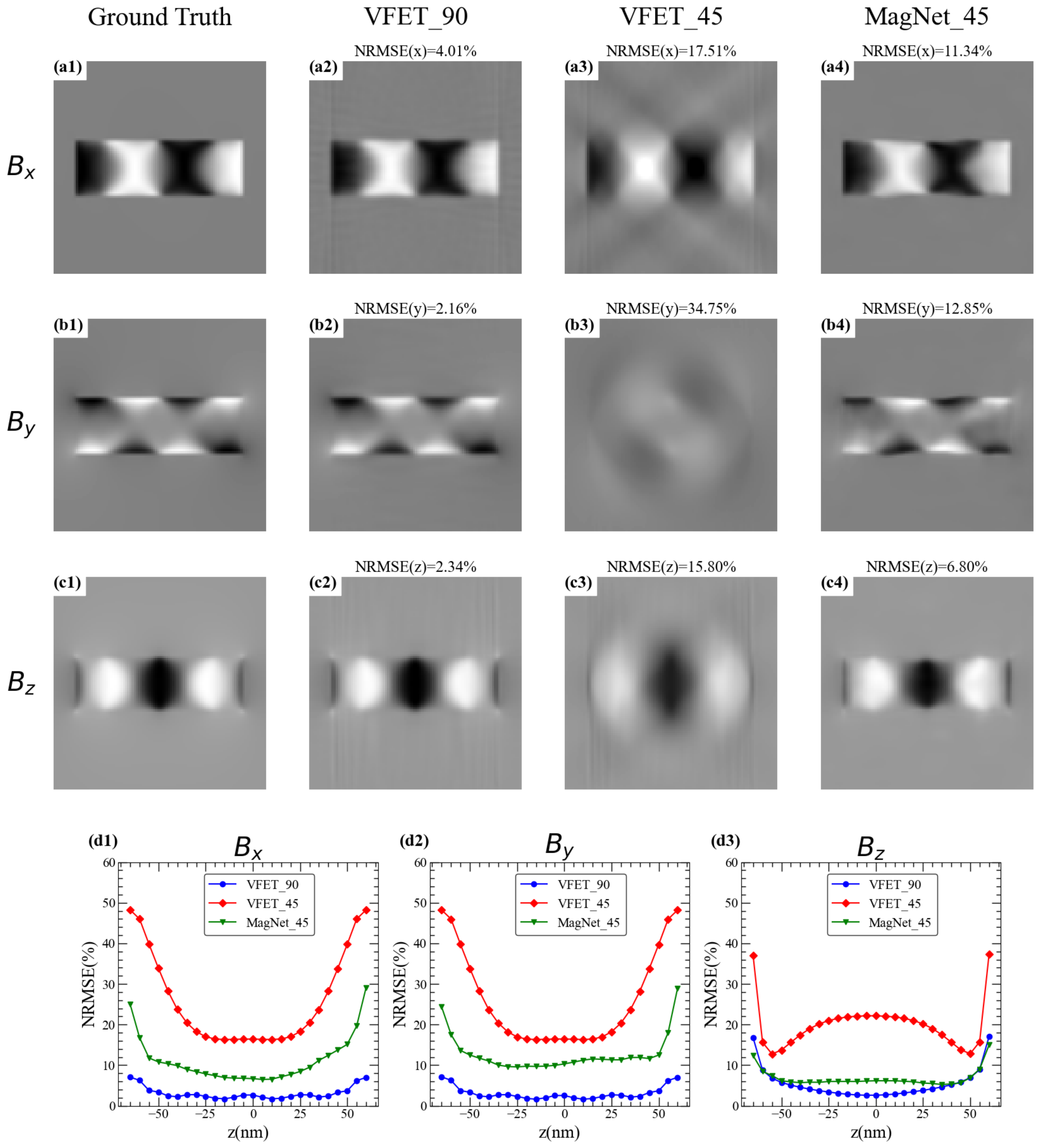}
	\caption{$x=0$ slice of reconstruction results from noise-free phases.
		The ground truth, VFET\_90, VFET\_45, and MagNet\_45 are shown in the first to the fourth column.
        (d1)-(d3) Single-layer NRMSE of $B_x$, $B_y$ and $B_z$ with different $z$ coordinate.}
	\label{profile_x}
\end{figure*}
%

\begin{figure*}[htb]
	\includegraphics[width=0.8\textwidth]{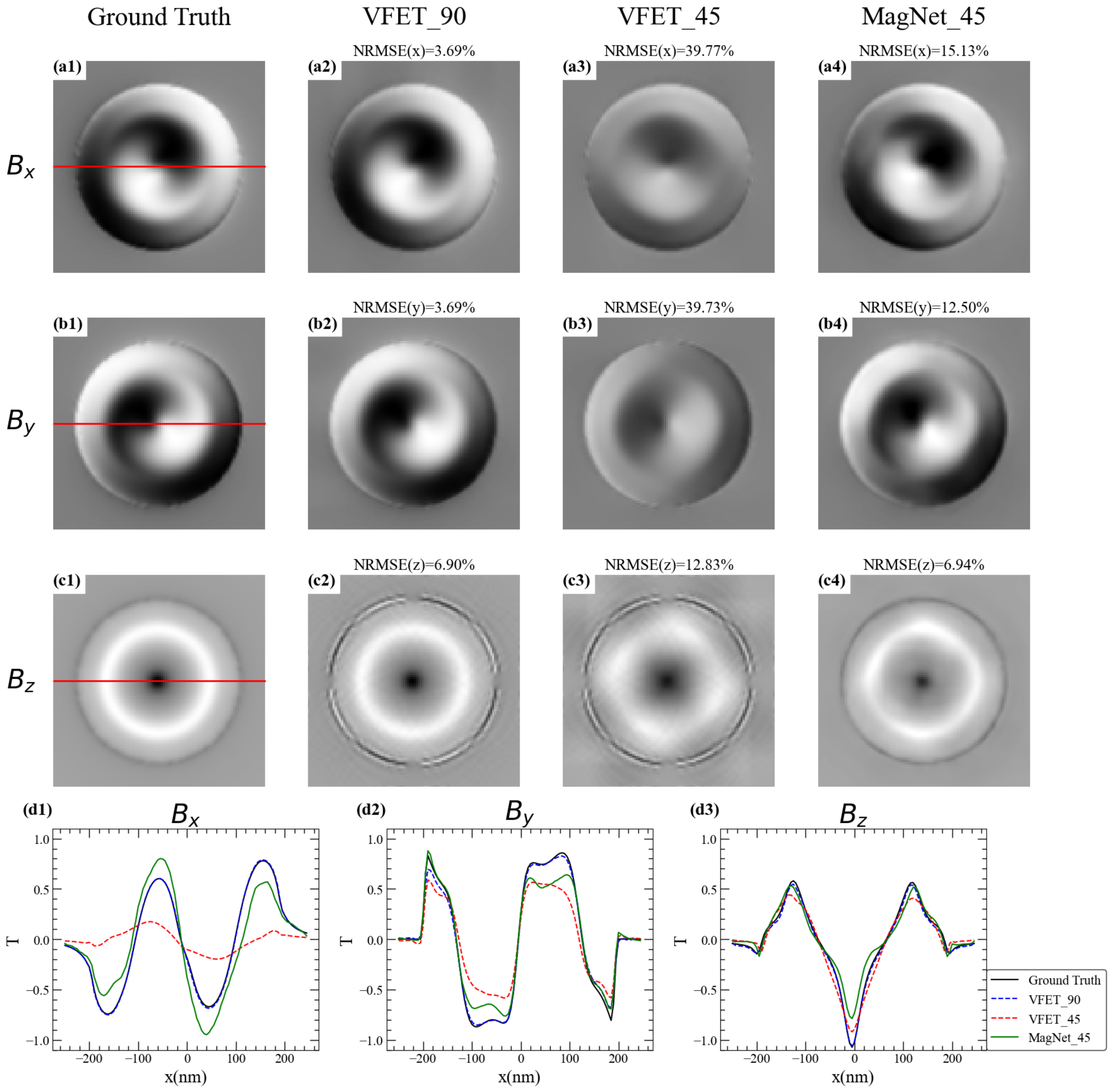}
	\caption{(a-c) $z=50nm$ slice of the ground truth, VFET\_90, VFET\_45 and MagNet\_45 .
		(d) $y=0$ line plot of each component.}
	\label{mw1_surface}
\end{figure*}
We use a skyrmion sample as a demo to show the comparison between reconstruction results from MagNet and VFET.
Figure \ref{mag} shows the upper and bottom surfaces, as well as the $x=0$ profile of the 3D magnetic texture. 
The skyrmion center is along the $z-$ direction.
Similar to the simulation in the previous work\cite{montoya2017tailoring}, the domain wall is Bloch-like in the central layers and becomes more N\'eel-like as approaching surfaces.
The diameter and thickness of the disc are 400nm and 130nm, respectively.
Figure \ref{profile_x} shows the reconstruction results on $x=0$ section.
The ground truth, VFET\_90 (complete tilt series), VFET\_45 and MagNet\_45 are shown in the first to the fourth columns, and $B_x$, $B_y$ and $B_z$ components are shown in the first to the third rows, respectively.
Each figure is titled with the NRMSE of corresponding component at x=0 layer.
Figures of the same row share a same color scale with the ground-truth image.
It shows that all three components of VFET\_45 suffer a blurred boundary caused by the missing wedge,
while the boundary of MagNet\_45 is sharp and clean.
%
The NRMSEs of the three components at x=0 layer are reduced from $17.51\%$, $34.75\%$, and $15.80\%$ to $11.34\%$, $12.85\%$, and $6.80\%$, respectively.
Figure \ref{profile_x}(d1)-(d3) show the single-layer NRMSE of three components at different z coordinates.
It shows that the NRMSE improvement of MagNet\_45 compared with VFET\_45 is about 5\% to 30\% at different layers.
Figure \ref{mw1_surface} shows the reconstructed results and the ground truth on $z=50nm$ layer.
It shows that the N\'eel cap is reconstructed by VFET\_90, but is not reconstructed by VFET\_45.
This indicates that the information that constructs the N\'eel cap is mostly within the missing wedge.
With the help of MagNet, the N\'eel cap can be reconstructed.
The NRMSEs of $B_x$ and $B_y$ are significantly reduced from $39.77\%$ and $39.73\%$ to $15.13\%$ and $12.50\%$.
Figure \ref{mw1_surface}(d) shows a comparison of reconstruction results from different models and ground-truth, indicating obvious enhancements on the $B_x$ and $B_y$ components.
\begin{figure*}[htb]
	\includegraphics[width=0.7\textwidth]{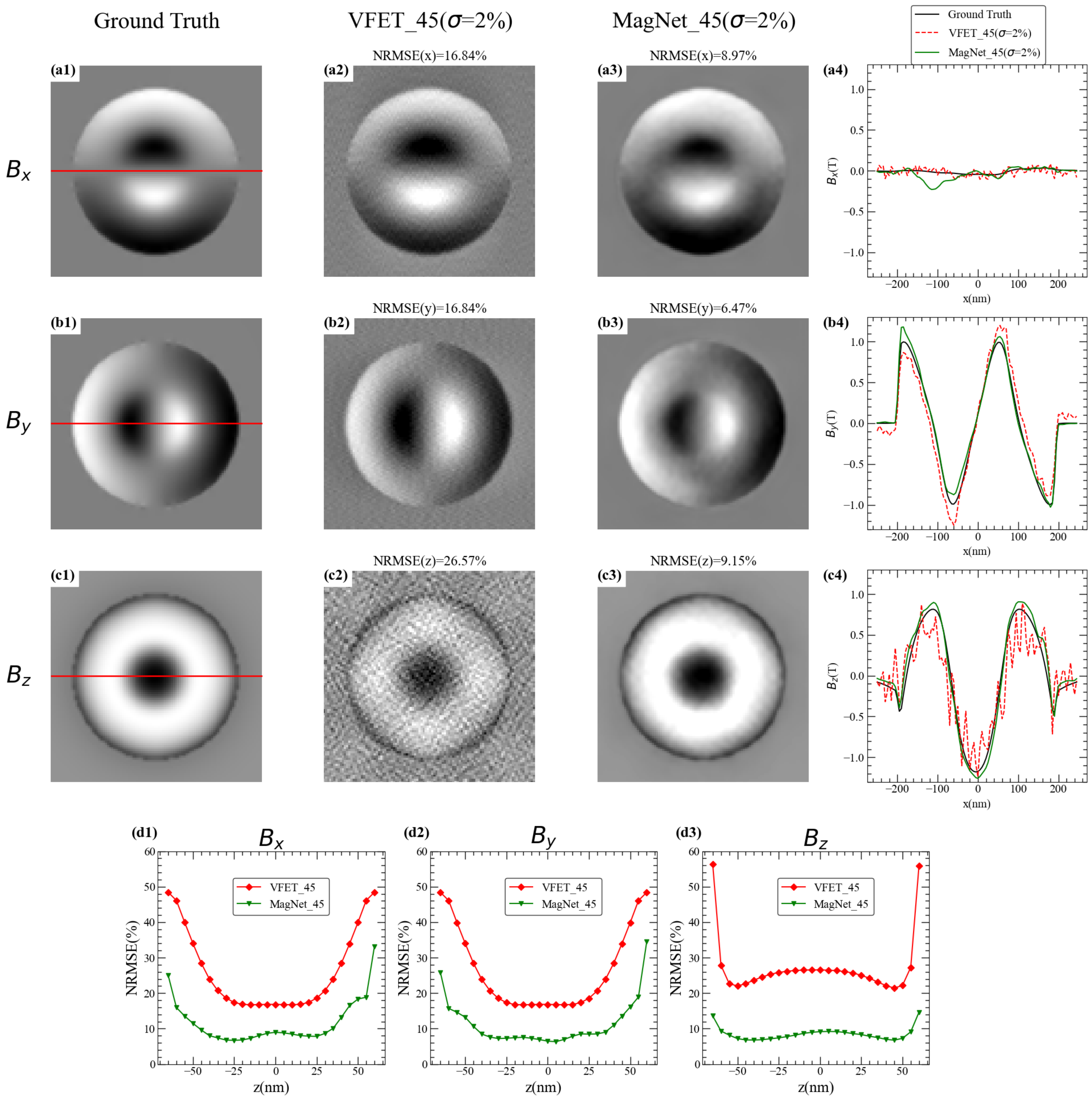}
	\caption{$z=0$ slice of the reconstruction results from $2\%$ Gaussian noise phases.
		The ground truth, VFET\_45 and MagNet\_45 are shown in the first to the third column.
		The fourth column shows the $y=0$ line plot.
		(d1)-(d3) show the single-layer NRMSE plots of $B_x$,$B_y$ and $B_z$, respectively. }
	\label{noise1}
\end{figure*}
\begin{figure*}[htb]
	\includegraphics[width=0.9\textwidth]{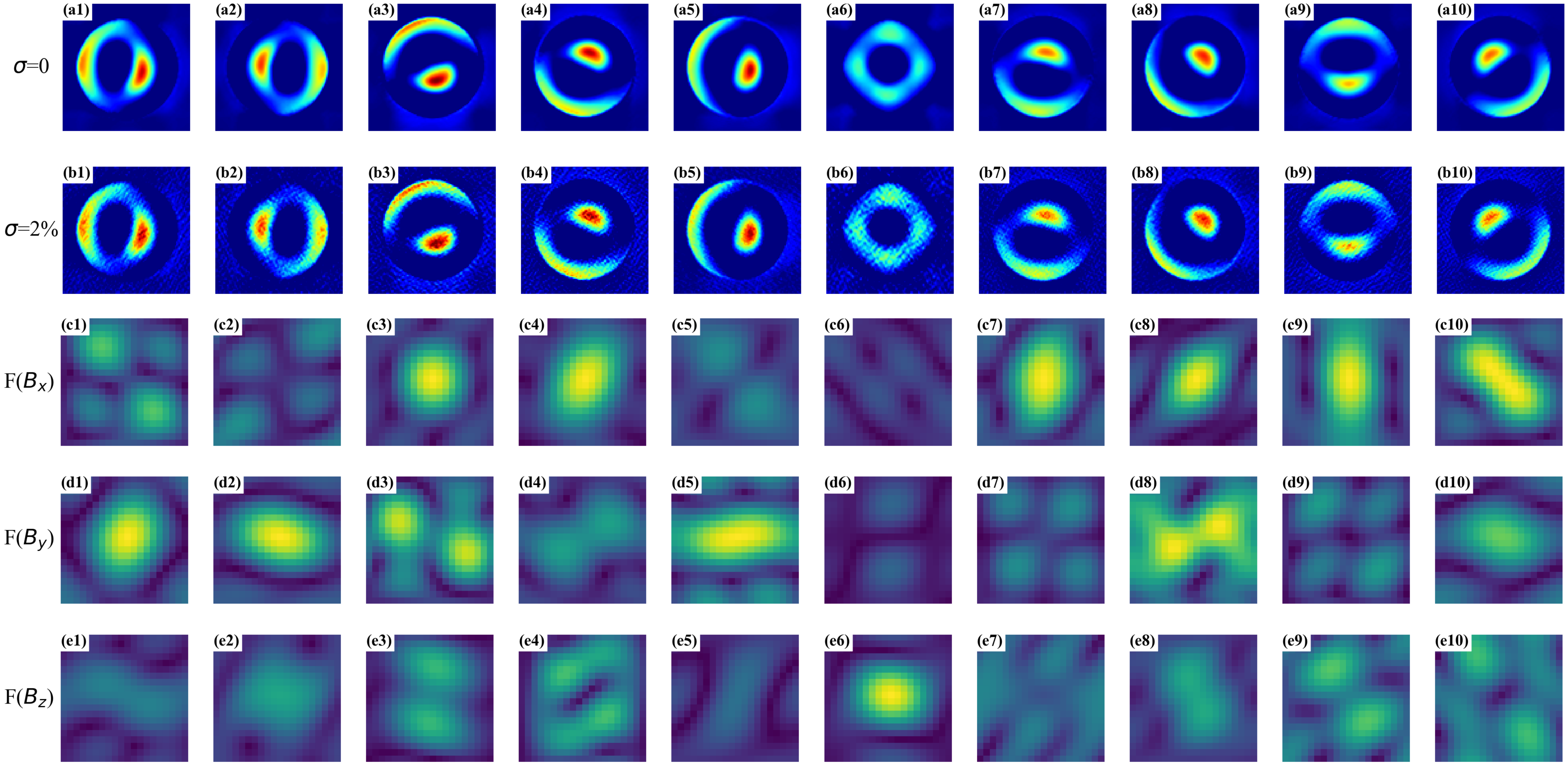}
	\caption{$z=0$ slice of top 10 strongest output features of the first CL of MagNet from (a) noise-free and (b) $2\%$ Gaussian noise input.
		(c)-(e) $k_z=0$ slice of Fourier intensity of corresponding convolution kernel from $B_x$, $B_y$ and $B_z$ input channel, respectively.
		(a1-10) and (b1-10) share a same color scale. Figures belong to a same column in (c)-(e) share a same color scale.
	}
	\label{features}
\end{figure*}

Figure \ref{noise1} shows the reconstruction results when the phases are contaminated by $2\%$ Gaussian noise.
The $z=0$ slice of the ground truth, VFET\_45 and MagNet\_45 are shown in the first to the third column, respectively.
The $y=z=0$ line profiles are shown in the fourth column.
%
%
%
The NRMSEs of $B_x$ and $B_y$ are both $16.84\%$, and the NRMSE of $B_z$ is $26.57\%$.
With the presence of noise, MagNet still keeps a stable performance,
and the NRMSE of the three components are only $8.97\%$, $6.47\%$, and $9.15\%$.
Figure \ref{noise1}(d1-d3) show the single-layer NRMSE of the three components.
Compared with its counterpart in the noise-free situation, the NRMSE of VFET\_45 of $B_z$ component has much more increase than that of $B_x$ and $B_y$.
On the other hand, the NRMSE increase of MagNet is tiny, which indicates the noise robustness of MagNet.
%
%

%
When $B_z$ is calculated by the gradient operator $\nabla \cdot \mathbf{B} =0$ which is a high-pass filter in x and y directions, 
the reconstruction errors of $B_x$ and $B_y$ are linearly amplified with the frequency in x and y directions, respectively.
In real experiments, the noise amplification could be suppressed by applying low-pass filters such as Butterworth-type filter~\cite{wolf2022unveiling}.
However, the choice of filter function is empirical, and different filters could lead to very different results.
Moreover, k-space filters are also required to reconstruct $B_x$ and $B_y$. 
If the relationship between components is considered, the choice of filter for each component would be rather complicated.
On the other hand, the convolutional layer(CL) of MagNet can be regarded as several multi-channel k-space filters.
%
%
%
For a CL with input channel number $m$, the tensor in the $j-th$ output channel is:
\begin{equation}
	y_j = \sum_{i=0,1...m} x_i * K_{ij}
\end{equation}
where $x_i$ and $y_j$ donates the $i-th$ input channel and the $j-th$ output channel respectively, $*$ donates the convolution operator, and $K_{ij}$ is the convolution kernel between those two channels.
In other words, every output channel accepts information from all input channels.
For example, the first CL of MagNet accepts $B_x$, $B_y$ and $B_z$ as input, so the 64 output features correspond to 64 different combinations of filtered $B_x$, $B_y$ and $B_z$.
Figure \ref{features}(a) and (b) show the $z=0$ slice of top 10 most strongly activated features of MagNet's first CL with ReLU activation function from noise-free input and noisy input.
$k_z= 0$ slice of Fourier amplitude of corresponding convolution kernel are shown in (c)-(e).
%
%
Taking the strongest output feature as an instance, the Fourier amplitude of $B_y$ (shown in (d1)) is much stronger than that of $B_x$ (shown in (c1)) and $B_z$(shown in (e1)).
As a result, the activated feature is mainly extracted from $B_y$ component.
This complex channel interaction provides rich prior knowledge of the relation between different input components.
%
%

\maketitle	
\section{Conclusion}	
\label{conclusion}

In conclusion, we proposed MagNet, a deep-learning enhanced VFET method, to reduce the reconstruction error caused by missing wedge problem.
The reconstruction quality for all testing samples are significantly improved compared with the conventional method when the maximum tile angle is below $\pm 60 \degree$.
%
%
%
A Bloch-type skyrmion with surface modulation is successfully reconstructed with phase shifts gathered at $\pm 45 \degree$ tilt range.
The reconstruction of the given example remains stable in the presence of $2 \%$ Gaussian noise.
It shows that MagNet is promising for real experimental applications.


To improve the performance of data-driven VFET, establishing a larger data library with classified magnetic structures is of vital importance. 
%
A widely accepted library can be used not only for training data-driven reconstruction methods, but also for evaluating the performance of reconstruction methods.
During the real-world applications of deep-learning VFET algorithms, users can add specific magnetic structures 
into the library. 
Finally, although we displayed the application with only one conventional algorithm, MagNet has the potential to work with any other conventional algorithms of VFET. 
With proper architecture design, it could also work together with modern X-ray reconstruction algorithms. 

\maketitle	
\section*{Data availability}	
Data will be made available on request.

\maketitle	
\section*{acknowledge}	
We are immensely grateful to Rafal E. Dunin-Borkowski and Fengshan Zheng for providing insight and expertise that greatly facilitate the research.
We would also like to thank Daniel Wolf for sharing his experience from his previous work with us during the course of this research.
This work was supported by the Office of Basic Energy Sciences, Division of Materials Sciences and Engineering, U.S. Department of Energy, under Award No. DE-SC0020221;
H.D. was supported by the Strategic Priority Research Program of Chinese Academy of Sciences, Grant No. XDB33030100, and the Equipment Development Project of Chinese Academy of Sciences,Grant No. YJKYYQ20180012.

\bibliographystyle{apsrev4-1}
\bibliography{references}
\end{document}